\documentclass[11pt]{article}
\usepackage{vietnam,epsfig}

\def\ao{{\em App. Optics.}}
\def\aap{{\em Astronomy \& Astrophysics}}
\def\apjl{{\em Ap.~J.~Lett.}}
\def\apj{{\em Ap.~J.}}
\def\ASTROPH{{\em astro-ph/}}

\def\be{\begin{equation}}
\def\ee{\end{equation}}
\def\bea{\begin{eqnarray}}
\def\eea{\end{eqnarray}}

\begin{document}
\vspace*{3cm}
\title{CMB FROM THE SOUTH POLE: PAST, PRESENT, AND FUTURE}

\author{ J.M. KOVAC and D. BARKATS }

\address{California Institute of Technology, Department of Physics, \\
1200 E California Bd, MC 59-33, Pasadena, CA 91125}

\maketitle\abstracts{
South Pole Station offers a unique combination of high, dry, stable conditions and
well-developed support facilities.
Over the past 20 years, a sequence of increasingly sophisticated CMB experiments at Pole
have built on the experience of early pioneering efforts, producing a number
of landmark contributions to the field.
Telescopes at the South Pole were
among the first to make repeated detections of degree-scale CMB
temperature anisotropy and to map out the harmonic structure
of its acoustic peaks.  More recent achievements include
the first detection of polarization of the CMB
and the most precise measurements of the temperature power spectrum at small angular scales.
New CMB telescopes at the South Pole are now
making ultra-deep observations of the large-scale polarization of
the CMB and of its secondary temperature anisotropies on arcminute scales.
These two observing goals represent the current frontiers of CMB research,
focused on constraining Inflation and the nature of Dark Energy.
The South Pole now hosts an array of CMB observing
platforms covering a wide range of angular scales and
supporting very long integration times on the cleanest sky available, and thus
should play an increasing 
role in pushing these frontiers of CMB research. 
}

\section{Introduction}


The bottom of the world is surprisingly well represented at
this 6th Rencontres du Vietnam 2006, with
four separate experiments sited at the U.S. National Science Foundation's South Pole Station reporting
science results in the parallel sessions:
one neutrino telescope, IceCube (K. Hoffman),
and three telescopes which measure
different aspects of the Cosmic Microwave Background,
QUAD (K.~Ganga), BICEP (C.~D.~Dowell), and ACBAR (C.~Reichardt).
The prominence of the latter here
reflects the increasingly important contribution the South Pole is making to CMB studies,
and by extension to our understanding of cosmology.

As the initiation of the International Polar Year (IPY) 2007-2008 turns a spotlight
of attention toward scientific efforts in Antarctica, it is perhaps timely to review the
history of efforts to measure CMB from the South Pole, to discuss some of
the unique characteristics of the site, and describe the directions that current and
future efforts there are taking.
An excellent overview of the history of astrophysics in Antarctica, including
efforts in cosmic ray, IR, sub-millimeter, and neutrino astronomy, is provided
by Indermuehle et al~\cite{2005PASA...22...73I}; here we concentrate just
on the CMB, which in recent years has become the major focus of photon astronomy
at the South Pole.


\section{Site Characteristics}

Isolated in the middle of the Antarctic plateau, the South Pole is a unique
site for observations in the millimeter and sub-millimeter windows. The
site
combines three 
characteristics necessary for high transmission: it is high (with an average
pressure altitude of 3200~m or 681~mbar), dry (less than 0.5~mm of precipitable
water vapor over half of the time), and cold (average annual: -49~C, minimum:
-82~C).\cite{1998ASPC..141..289L}
Although it shares some of these attributes with other mm/submm sites
(Mauna Kea, Hawai or Atacama desert, Chile), 
site surveys suggest that the South Pole
provides better consistency of mm transmission.\cite{rad2000}
Very small daily thermal variations and wind patterns dominated by
katabatic flow make the atmosphere overhead extremely stable.\cite{cara}
This aspect is extremely important for CMB experiments;
median wintertime fluctuations at 150~GHz have been found to be 30 times lower than
at the ALMA test site in Atacama.\cite{2005ApJ...662.1343B}

Because the sun rises and sets once per year,
sun contamination is absent for the six month winter.
Target observation fields remain at the same elevation in the
sky. They do not set and integration on the field is therefore limited only by the
experiment's operational efficiency rather than by the field's availability.

\begin{figure}
\begin{center}
\includegraphics[height=2.3in]{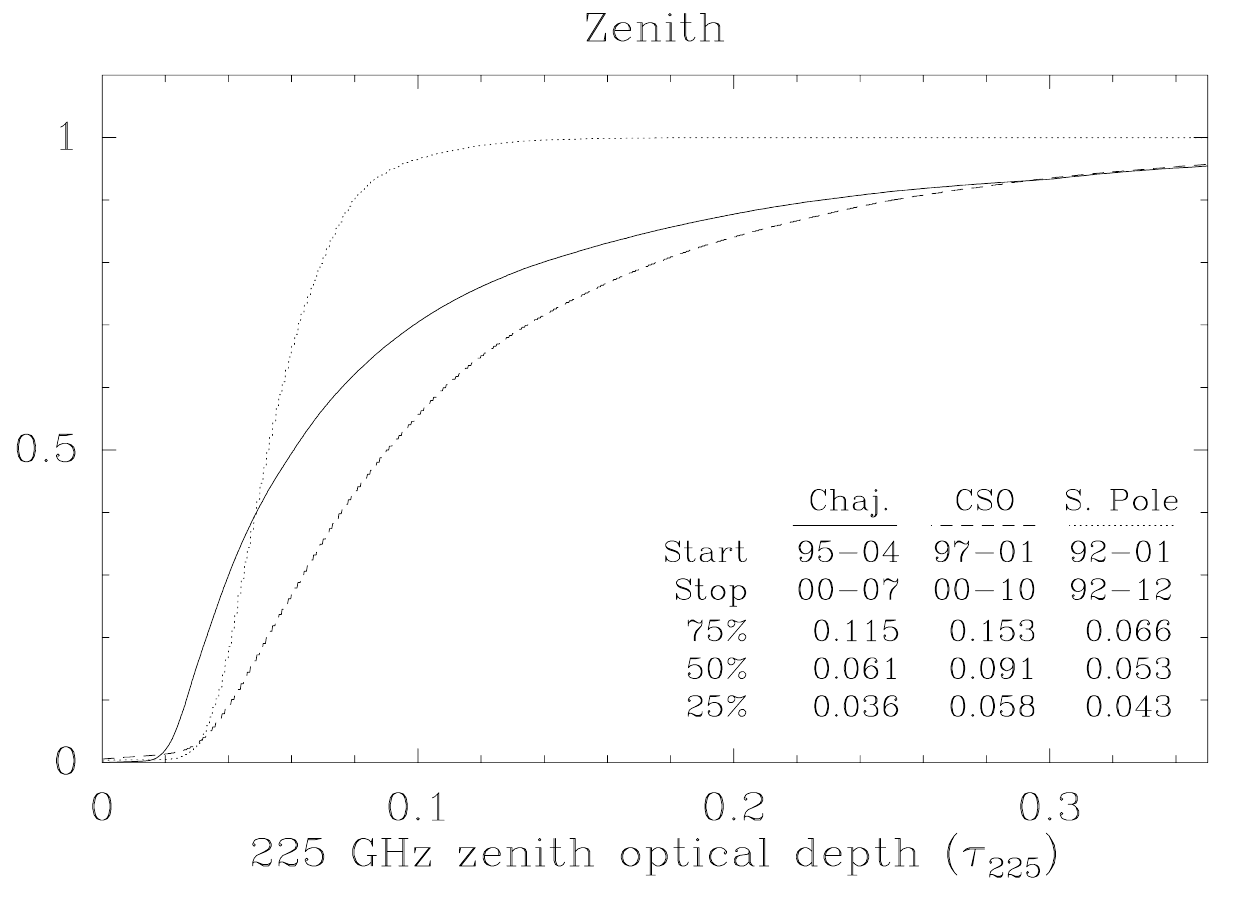}
\includegraphics[height=2.3in]{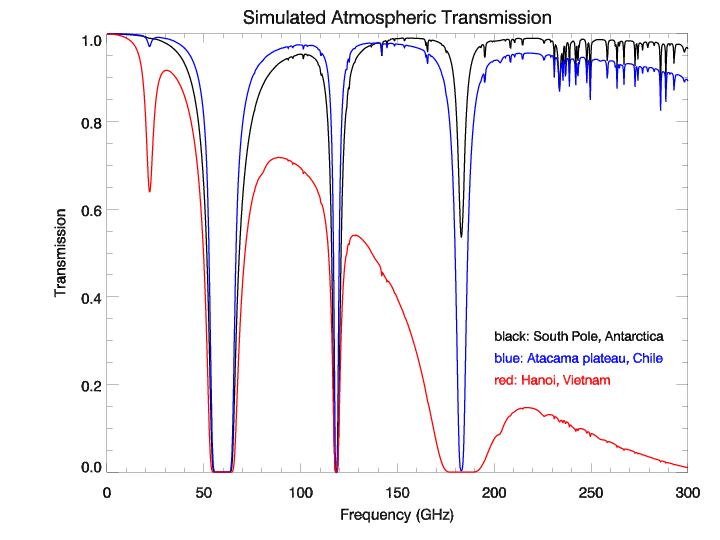}
\end{center}
\addtocounter{figure}{1}
\footnotesize
Figure 1:
Left: Comparison of 225 GHz opacity, which is dominated by water vapor, between
  Chajnantor (ALMA site, Atacama, Chile), the CSO (Mauna Kea, Hawaii), 
  and the South Pole.\cite{rad2000}
  The best days in Chile are drier, but the South Pole enjoys a larger fraction
  of dry days.
Right: Atmospheric transmission model using the AT software~\cite{at} for typical
  winter conditions~\cite{1998ASPC..141..289L} at the South Pole (black)
  and the Atacama plateau (blue),
  and at a sea-level site, Hanoi, Vietnam (red, PWV= 60~mm). The 22 and
  180~GHz H$_2$O  line and 60 and 120~GHz O$_2$ line clearly
  delineate observing windows at $\sim$ 30, 100, 150, and 220 GHz.
\label{fig:pwv}
\end{figure}

\vspace{-0.1in}
\paragraph{Infrastructure and Logistics}

Through 
50 years of operation (since summer 1956-1957)
the Amundsen-Scott South Pole Station has developed an 
outstanding infrastructure capable of 
supporting all kinds of scientific experiments, including the peculiar
requirements 
of CMB experiments: transportation, communications, construction
support, electrical power, technical support, laboratory space,
accommodations, cryogenic support, to name only a few.

All cargo and personnel arrive at the South Pole in LC-130 Hercules.
Flights are restricted to a brief summer period, November through mid-February,
when temperatures permit the planes to land;
each summer sees approximately 300 flights.
%
The nine months of winter inaccessibility enforces a strict project discipline
and requires careful planning of the experiment for a whole year. 
During the austral winter, at most one or two team members --winter-overs-- 
stay behind to run
the experiment, freeing the rest of the team to concentrate on analyzing
incoming data. 

In January 2008 the NSF will dedicate the new South Pole station after 
a decade of major upgrades to science support and living facilities
at Pole.  For CMB experiments, these facilities already provide expanded observatory
space, a year-round liquid helium supply, and 80 GB/day of satellite data
transmission, transforming the station into a world-class observatory.

\section{The Heroic Age: 1984-1992}
\vspace{-0.10in}

\begin{figure}[t]
\begin{center}
\includegraphics[height=2.64in]{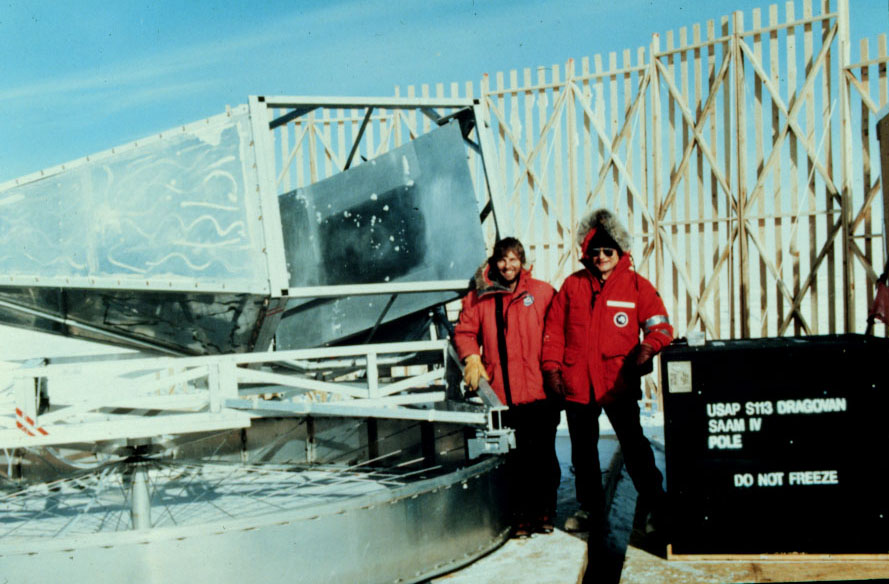}
\includegraphics[height=2.64in]{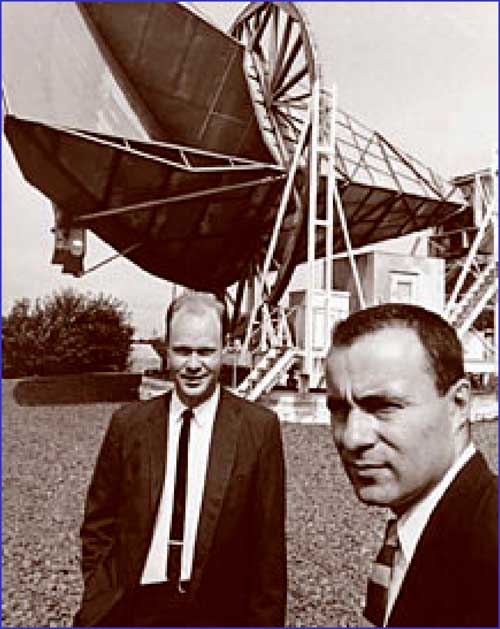}
\includegraphics[height=1.46in]{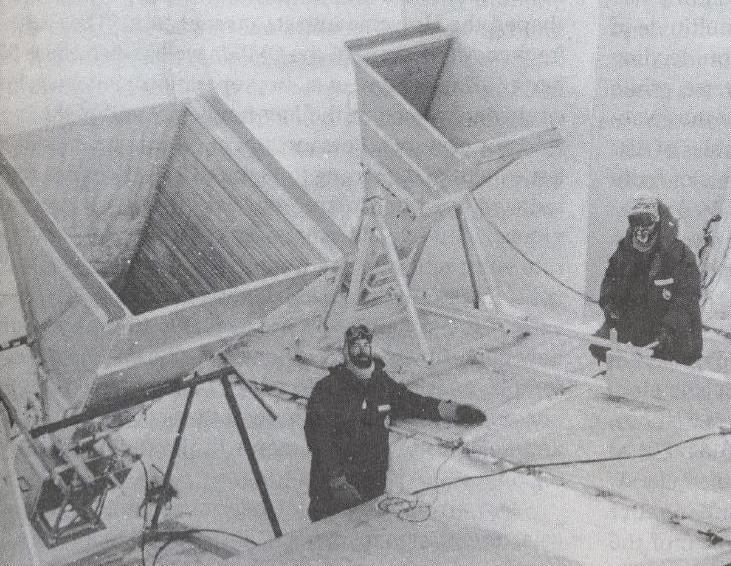}
\includegraphics[height=1.46in]{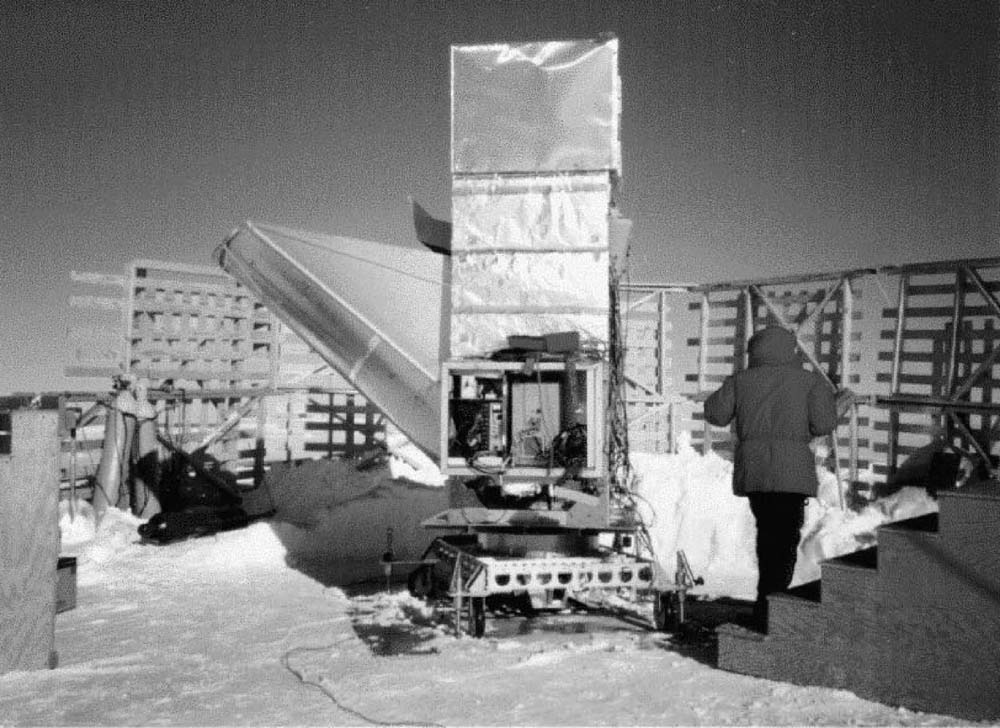}
\includegraphics[height=1.46in]{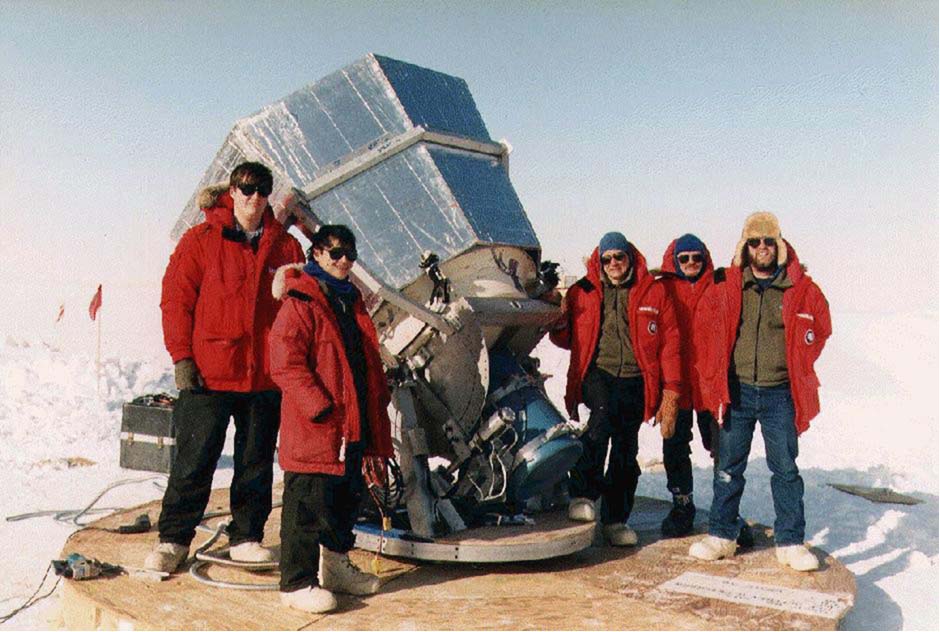}
\end{center}
\vspace{-0.1in}
\caption{
The Bell/Princeton telescope in 1986 was the first CMB telescope
fielded at Pole; it is seen here with M. Dragovan and R. Pernic in 1988 after
installation of the ``bicycle wheel'' azimuth track (top left);
the legacy of the original Bell Labs horn antenna with which Penzias and Wilson
detected CMB in 1965 (top right) is apparent.
Lower left: Two of the Smoot group's total power radiometers at
1.47 and 2~GHz inside a 4~m deep pit in the ice to serve as
ground and sun screens in summer 1991.
Lower middle: The UCSB group's ACME telescope, summer 1993.
Lower right: Princeton's 1.4m White Dish, H. Nguyen in the foreground and J. Peterson
at right, Jan 1993.
}
\label{fig:miss_piggy}
\vspace{-0.15in}
\end{figure}

Minimal facilities greeted the first experiment to take advantage of the
low sub-millimeter opacity at the South Pole.  This was a US-France collaboration, the
EMILIE (Emission Millimetrique) experiment,\cite{1986A&A...154...55P} mounted during
the 84-85 austral summer. With the help of M. Pomerantz, the French team
operated a 45-cm telescope at wavelengths near 900~$\mu$m to
measure the dust emission of the galactic center region. 
This experiment
provided a first test of the logistics that future CMB experiments would have
to cope with: liquid helium delivery, remote power, and heated lab space---provided initially by Jamesway tents.

The first effort to measure the anisotropy of the cosmic microwave background
came in the summer 86-87, led by Mark Dragovan and Tony Stark of ATT/Bell Labs
and Bob Pernic of the Yerkes Observatory, and again helped by Pomerantz.
This was a $1.2$m off-axis horn
antenna, initially operated with a single 400~GHz
bolometer.\cite{1990ApOpt..29..463D}  The telescope was at first only
steerable in elevation (Fig~\ref{fig:miss_piggy}) but was improved in
subsequent years to track in azimuth (thus the name ``bicycle wheel
experiment''). It was located at a site 1 mile grid south of the station,
in what would be known as ``CMBR Land''. This experiment confirmed the quality of the site
and paved the way for the increasingly sophisticated series 
of CMB experiments which followed.

Three other research groups joined the Bell/Princeton group in CMBR Land
during the 1988-1992 summers.
A Berkeley team led by George Smoot installed 6 radiometers
at 0.6, 0.8, 2.5, 3,75, 7.5, and 100~GHz
to try to improve 
previous measurements of the CMB
temperature 
spectrum~\cite{1991ApJ...381..341D,ber93}.
During two campaigns in summer 89-90 and 91-92, they
probed for long wavelength CMB spectrum distortion using total power or
Dicke-switched differential radiometers.

A UCSB team led by Phil Lubin
installed various receivers 
during their three South Pole 
campaigns (summers 88-89, 90-91, 93-94).
Their most successful results came from their last summer of
observation with a Ka and Q-band HEMT radiometer placed on a 1-m off-axis
Gregorian telescope, 
Analysis of these data eventually yielded a detection
of anisotropy at degree angular scales  ($\Delta T_{rms} = 41.2^{+15.5}_{-6.7}~\mu$K at $36 < \ell < 106$)
in a frequency range between 26 and 45~GHz, with a spectrum consistent
with a CMB thermal spectrum.\cite{1995ApJ...443L..57G}

A Princeton group led by Jeff Peterson 
fielded the White Dish experiment, a 1.4-m on-axis
telescope using a single-mode bolometer at 90~GHz cooled with an
ADR fridge.\cite{1993ApJ...419L..45T}
From their two summers of
observations (summer 91-92, 92-93), White Dish provided the tightest upper
limit on the CMB anisotropy at sub-degree angular scales ($\Delta T_{rms} <
62~\mu$K at $\ell \approx 800$).

\begin{figure}[t]
\begin{center}
\includegraphics[height=1.75in]{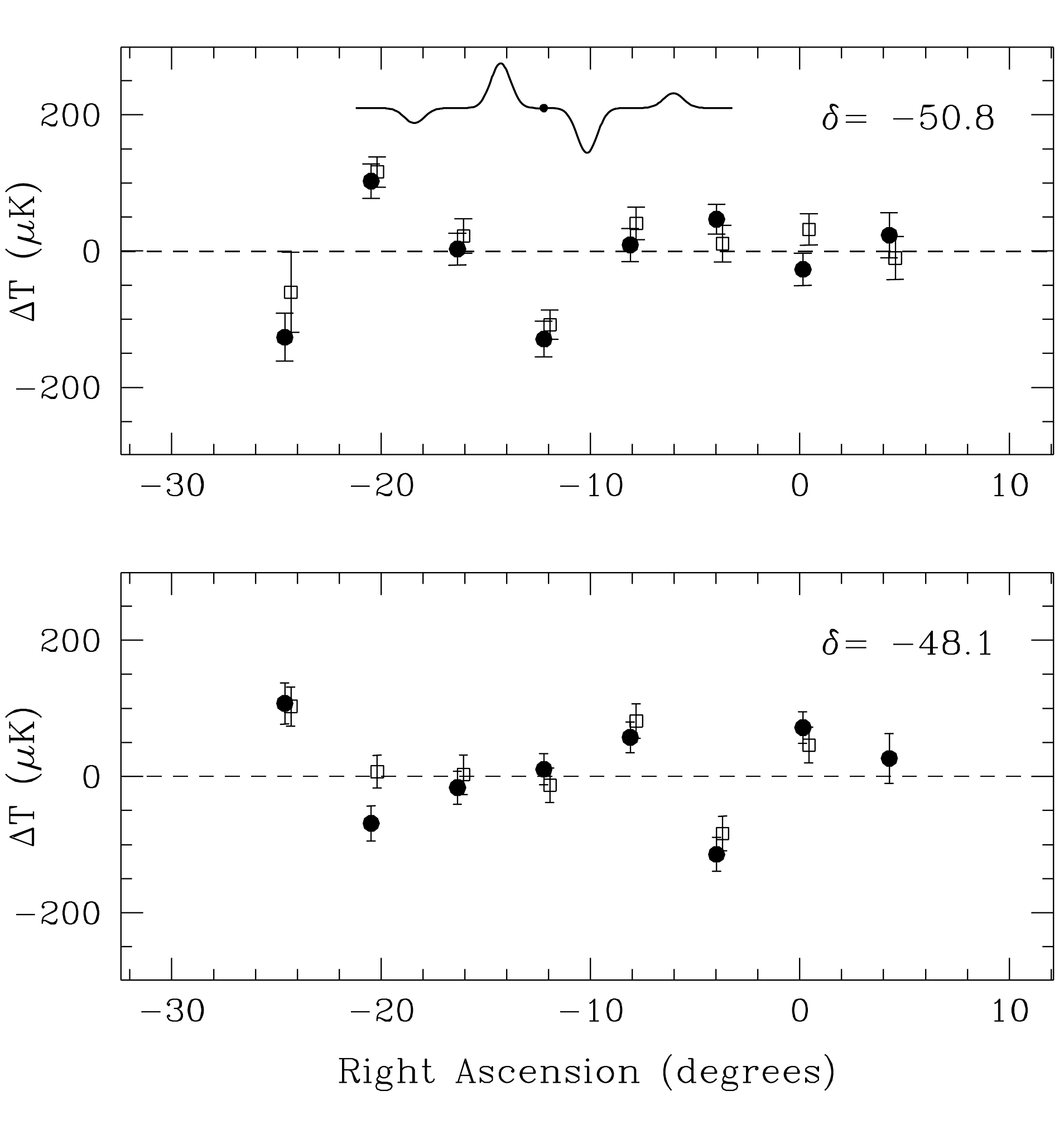}
\includegraphics[height=1.75in]{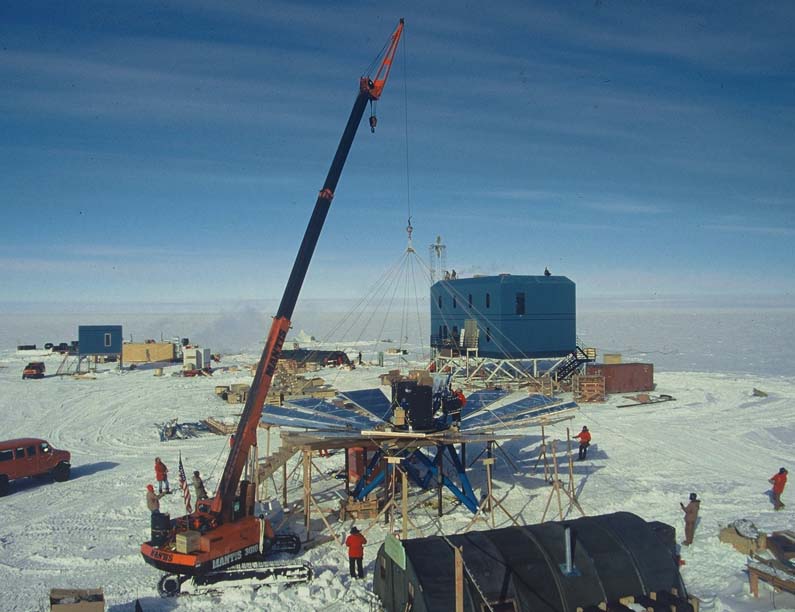}
\includegraphics[height=1.75in]{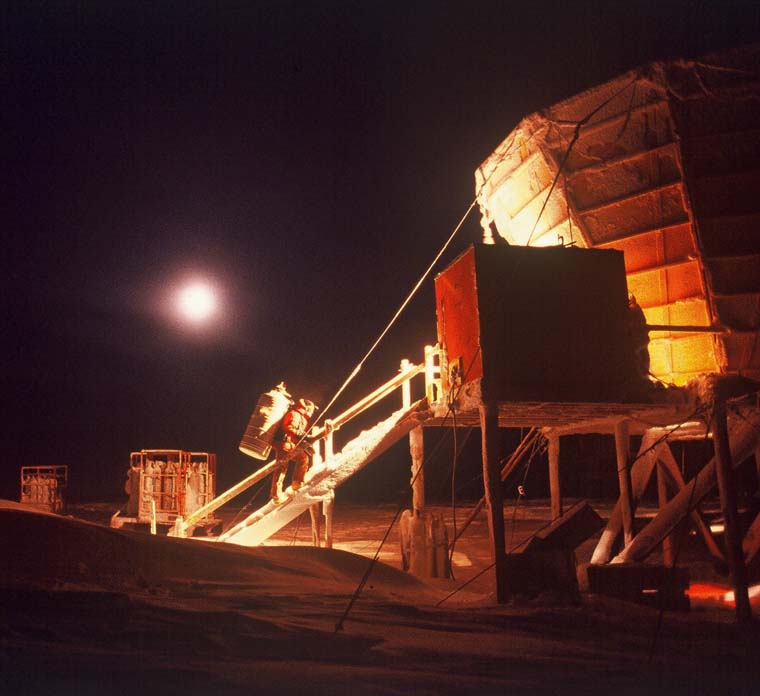}
\includegraphics[width=0.96\textwidth]{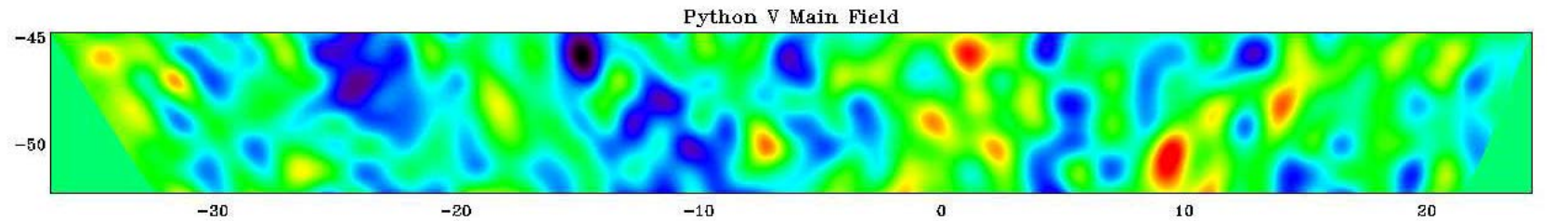}
\end{center}
\addtocounter{figure}{1}
\footnotesize
Figure 3:
The Python telescope detected CMB anisotropy on degree angular scales in 1992,
a few months after COBE results had revealed fluctuations on much larger scales.
In 1993 the telescope repeated and confirmed these observations (upper left)
before being relocated as the first permanently installed CMB telescope in
the new Dark Sector area organized by CARA (center).
Hard lessons learned during first winter operations with Python in 1994 (upper right)
guided the design of future telescope facilities.
In its final season 1996-97, Python produced a significant degree
scale temperature map of the CMB in the Southern Hole (lower). 
\label{fig:python}
\vspace{-0.15in}
\end{figure}

\vspace{-0.15in}
\section{Building the Dark Sector: 1992-2005}
\vspace{-0.10in}

1992 was a pivotal year for CMB studies at the Pole and elsewhere.  The initial detection
of CMB anisotropy at very large angular scales by the COBE satellite~\cite{1992ApJ...396L...1S}
moved the pace of discovery permanently into high gear.  In 1991 NSF had established
CARA, the Center for Astrophysical Research in Antarctica, to organize
IR, submm, and microwave observing facilities at the Pole in a new
``Dark Sector''.  CARA's first CMB telescope, Python, debuted at 
Pole in late 1992.  

\vspace{-0.15in}
\paragraph{Python:}
Led by M. Dragovan at Princeton, Python
was a 0.75~m off-axis telescope with a fast chopping primary flat.
It was first operated in late 1992
from old CMBR Land and quickly detected
CMB anisotropy on degree scales,\cite{1993AAS...182.7702D}
announcing results less than a year after COBE.  The next season Python
repeated these detections with multiple tests confirming the reproducibility of
the observed signal.\cite{1995ApJ...453L...1R}
That same summer, Python was relocated to a more permanent installation on a 
tower in the new Dark Sector, and in 1994 became the first CMB telescope
to operate in the winter at the South Pole.
Python's receiver was a state-of-the-art array of four 90~GHz bolometers cooled to 50~mK.
Its initial winter demonstrated the possibility of operating a complicated CMB experiment
through the long South Pole night, but also identified severe 
challenges.
Subsequent telescopes
incorporated lessons learned from Python
in the design of environmental enclosures, maintenance access, and cryogen facilities.
Python operated through the summer of 1996-97, eventually producing degree-scale maps
at both 90 and 40~GHz.\cite{2003ApJ...584..585C}


\begin{figure}
\begin{center}
\includegraphics[height=2.3in]{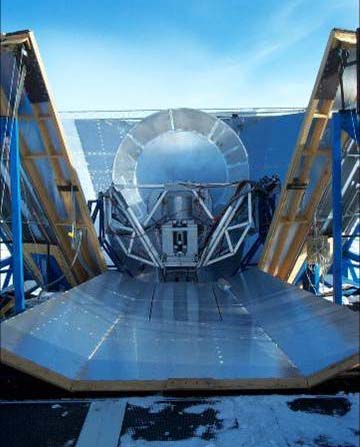}
\includegraphics[height=2.3in]{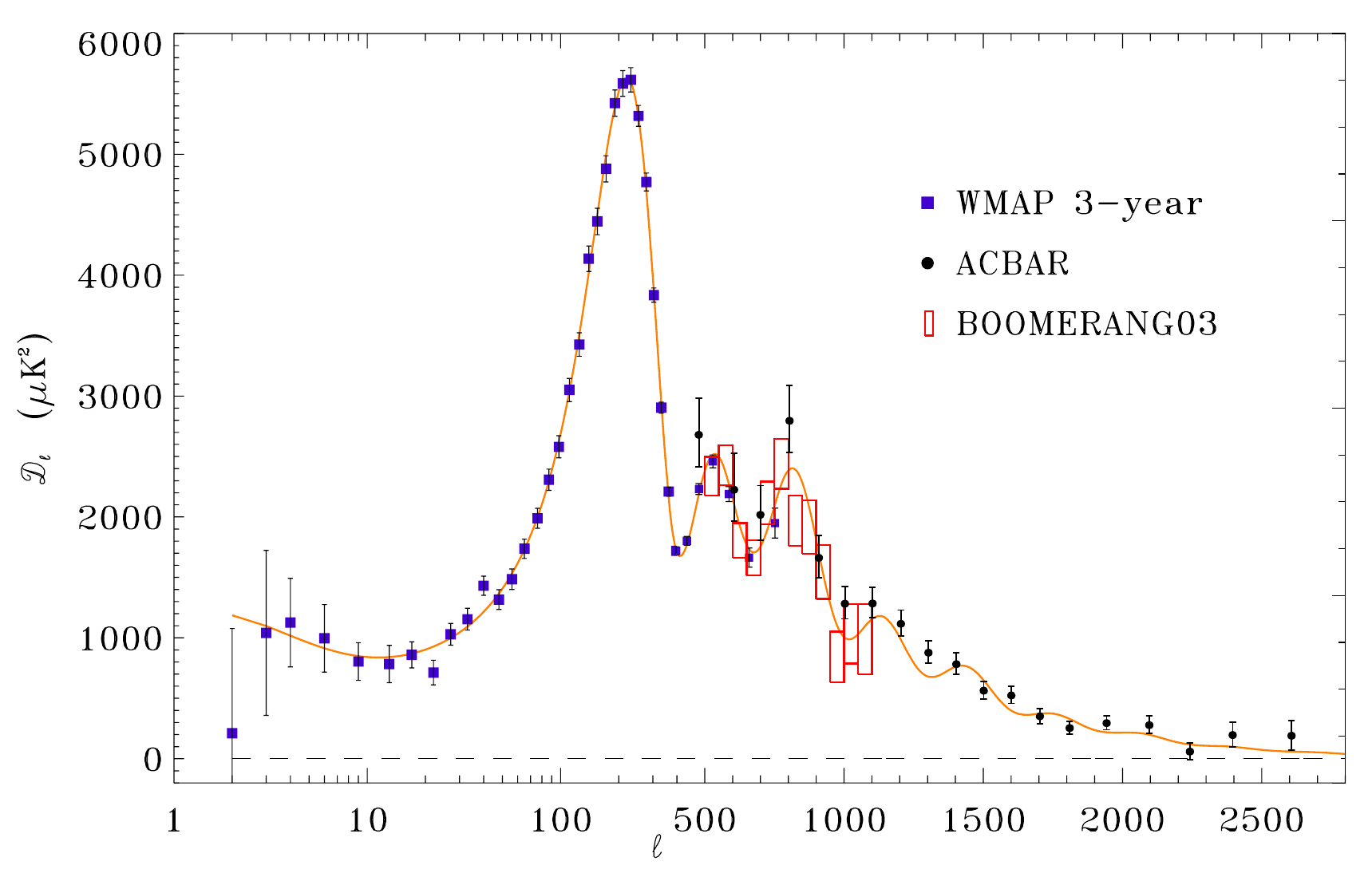}
\end{center}
\addtocounter{figure}{1}
\footnotesize
Figure 4:
The Viper telescope, a 2.1m off-axis Gregorian, was installed near the heated
control space of MAPO in 1998. Ground shield panels lower allowing
low-elevation calibrations (left).  The ACBAR receiver operated on Viper
2001-2005, producing high precision measurements at small angular scales
which have extended constraints on CMB power spectrum from the
largest scales, now well-measured by WMAP, through the damping tail (right).
\label{fig:acbar}
\vspace{-0.15in}
\end{figure}

\vspace{-0.1in}
\paragraph{VIPER/ACBAR:}
When the Martin A. Pomerantz Observatory (MAPO) was dedicated 1995, plans
were drawn to build a successor to Python into the new facility.
The Viper telescope, commissioned in January 1998,
was a 2.1~m off-axis Gregorian with a chopping tertiary, 
designed to provide larger throughput and higher angular resolution than Python.
The Arcminute Cosmology Bolometer Array Receiver, ACBAR,
harnessed the power of Viper with a 16 element bolometer array cooled to 250~mK.
It was first deployed on Viper in winter 2001 with a focal plane containing
150, 220, and 280~GHz pixels.  It was found that foreground confusion in clean
regions of the southern sky did not limit 150~GHz sensitivity,
and the number of 150~GHz pixels was increased to 8 for the 2002 winter,
and to all 16 pixels for ACBAR's final winter, 2005.
ACBAR results, reported in this meeting by C. Reichardt,
include extremely deep, high-resolution CMB temperature maps which
provide precise measurements of the CMB power spectrum at small scales,\cite{2006astro.ph.11198K}
and have been combined with results from CBI and WMAP to place the best current constraints
on cosmological parameters from the CMB.\cite{2006astro.ph..3449S}

\vspace{-0.1in}
\paragraph{DASI and QUAD:}
The Degree Angular Scale Interferometer was a compact 26-36 GHz interferometric array designed
to measure CMB temperature and polarization at angular scales 140 $< l <$ 910.  It was
installed on a tower adjacent to MAPO in late 1999 by a U. of Chicago team led by J. Carlstrom
and M. Dragovan,
and over the 2000 winter mapped 32 fields.
Results on the CMB temperature spectrum were published
in April 2001~\cite{2002ApJ...568...38H}
just over a year after data-taking commenced.
In a joint announcement with the Boomerang Antarctic balloon-borne experiment
it was revealed that both experiments had independently confirmed the 
harmonic peak structure of the temperature spectrum, and
in particular measured second and third peak amplitudes
consistent with predictions of BBN and dark matter.
The DASI receivers were upgraded with novel achromatic polarizers in early
2001 and polarized observations were conducted
over the following three winter seasons.
First results on CMB polarization were released in September
2002,\cite{2002Natur.420..772K}
and revealed at $5\sigma$ confidence
the first detection of CMB polarization.

\begin{figure}
\begin{center}
\includegraphics[height=1.65in]{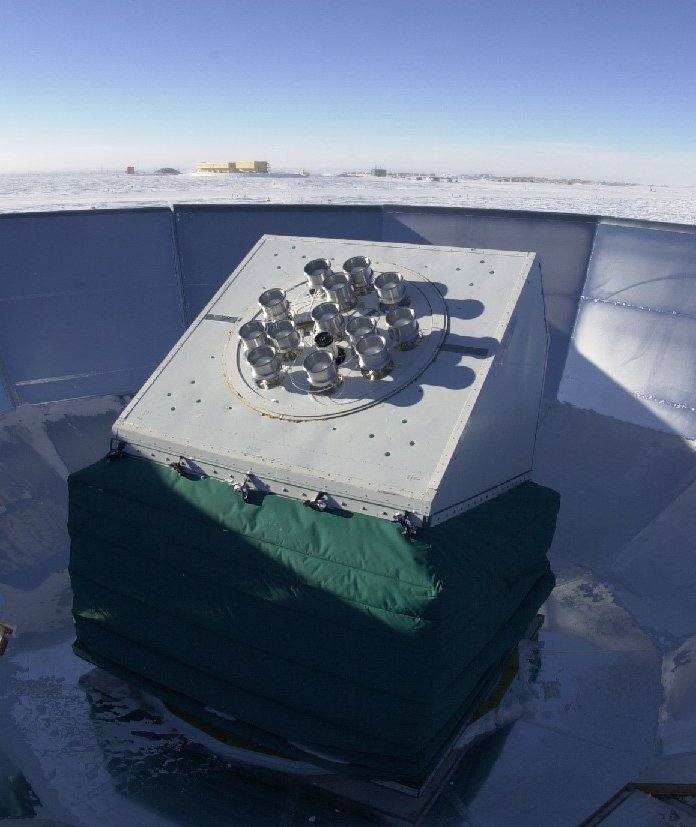}
\includegraphics[height=1.65in]{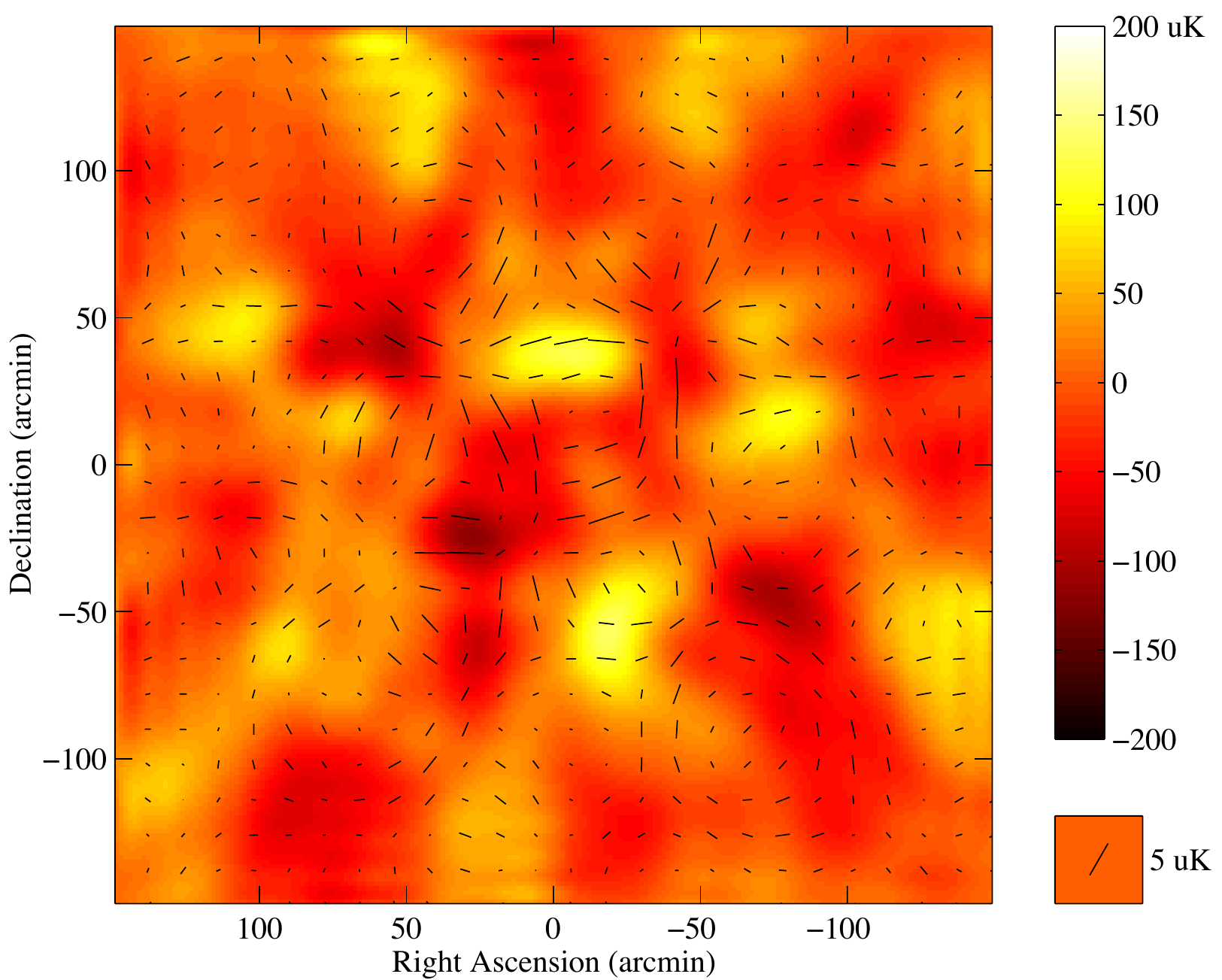}
\hspace{0.1in}
\includegraphics[height=1.65in]{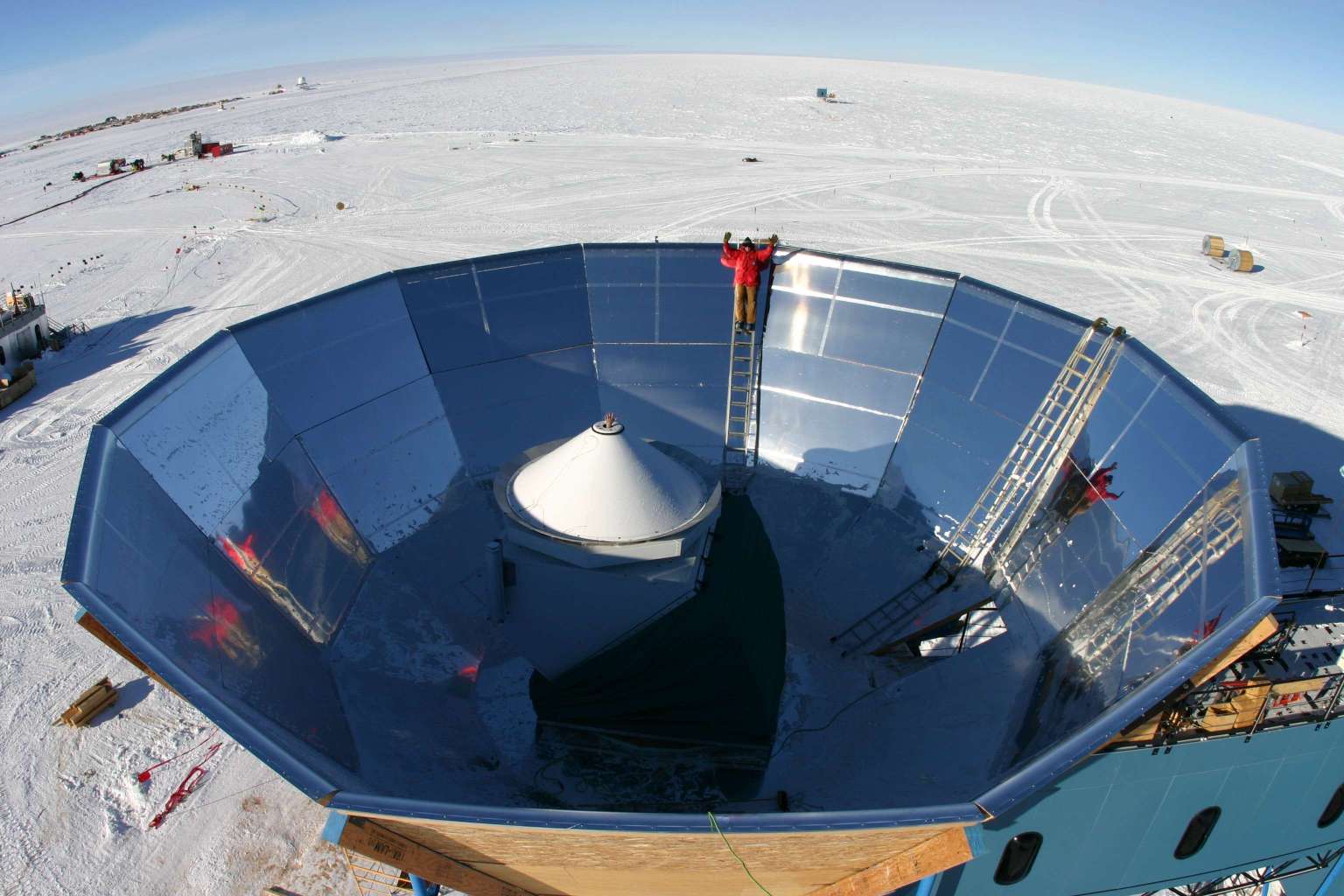}
\end{center}
\addtocounter{figure}{1}
\footnotesize
Figure 5:
DASI, a 13 element interferometer with a unique enclosed geometry, began mapping
the acoustic peaks of the CMB temperature spectrum in early 2000 (left).  The following
year polarization capability was added, resulting in the first detection of CMB
polarization, published in 2002 (center).  The QUAD experiment (right) began operation
in 2005, replacing the DASI array with a 2.6m Cassegrain telescope, a foam-cone supported
secondary, and a receiver housing 62 polarization-sensitive bolometers.
Now in its third observing season, QUAD is producing the deepest-yet maps of CMB E-mode
polarization at medium to small angular scales.
\label{fig:dasi_quad}
\end{figure}

Significant upgrades to DASI as an interferometer were unattractive due to the $n^2$ scaling of
the correlator, so a proposal was formed to mate the DASI platform, to be operated from Chicago
by C. Pryke, to the QUEST
2.6m Cassegrain telescope and receiver, under development by teams led by W. Gear at Cardiff
and S. Church at Stanford, respectively.  
QUAD was the result: a bolometric instrument on the DASI mount
boasting 62 polarization sensitive bolometers at 100 and 150~GHz.\cite{2006NewAR..50..984P}
QUAD is currently mapping E-mode polarization of the CMB from angular scales of 200 $< l <$ 2000,
as reported in this meeting by K. Ganga.

\section{New Challenges: 2005-future}

The search for the faint but unique signature of Inflation in the B-mode polarization of
the CMB at degree scales was identified by the 2005 Task Force on CMB Research as the
number one future priority for the field.  The second priority identified was the study of CMB
anisotropies on small scales, where SZ cluster and lensing surveys can track the
growth of structure and thus constrain properties of dark matter, dark energy, and neutrinos.

While there is no substitute for all-sky satellite missions for 
ultimate measurements of CMB power spectra, 
progress on these two new priorities in coming years is likely to be led from the ground.
The optimal strategy for \emph{discovery} of degree scale B-modes from Inflation is
extremely deep integration on a single $f_{\rm{sky}}\sim2$\% region, with foreground
avoidance a top priority.  Ground based telescopes, particularly those sited at the South Pole,
are ideal for observing such a region, see Figure 6. 
Arcminute scale CMB anisotropies can only be surveyed using large, ground based telescopes.
The new generation of CMB telescopes now operating from the South Pole are
targeted toward these two goals.

\begin{figure*}[t!] 
\begin{center}
\includegraphics[width=.53\textwidth]{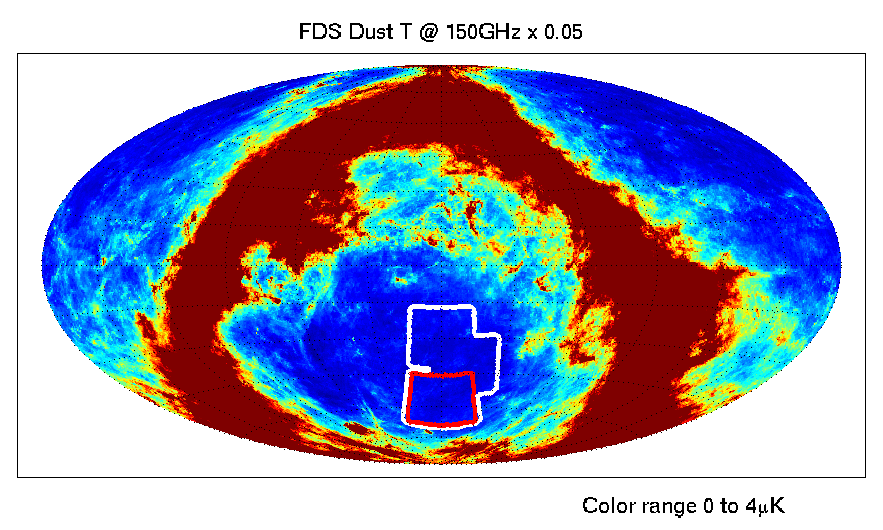}
\includegraphics[width=.45\textwidth]{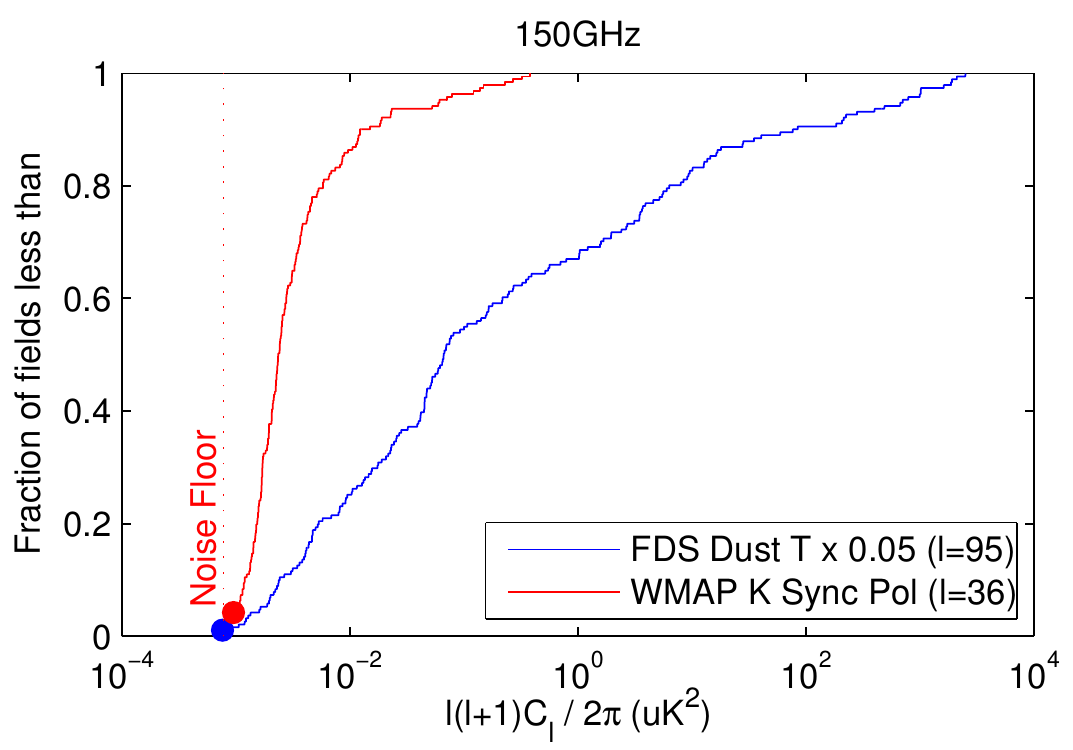}
\end{center}
\begin{flushright}
\footnotesize\emph{figure: C. Pryke/SPUD}
\end{flushright}
\addtocounter{figure}{1}
\footnotesize
Figure 6:
The ``Southern Hole'' is seen in the all-sky FDS model of thermal dust emission (left),
which is the dominant galactic foreground for CMB observations at high frequencies.
The red region is the 800 deg$^{2}$ ($f_{\rm{sky}}=2$\%) BICEP field; the white
boundary of the ``Hole'' is shown as the aggregate of all 800 deg$^{2}$ fields across the sky
with equal or lower dust power at $l$=95 (the few best fields in the north are still
slightly dustier).  On the right, assuming a 5\% polarization fraction, the polarized
contamination predicted from dust is compared to that from synchrotron, which dominates
at lower frequencies, for all 800 deg$^{2}$ fields (integral distribution) and for the
red field (dots).  
Note that the dust foreground exhibits greater variation than synchrotron, with
100x less dust power in the ``Southern Hole'' compared to typical a high galactic latitude
field at the median of the distribution.  Consequently, a very low minimum
in total foregrounds is expected at frequencies near 150~GHz (see
Figure 7). 
The South Pole site offers a continuous view of the Southern Hole
at high elevation; the Southern Hole is also visible from Atacama 
for up to six hours each day.
\label{fig:fore_maps}
\end{figure*}   

\begin{figure}[t]
\begin{center}
\includegraphics[height=2.2in]{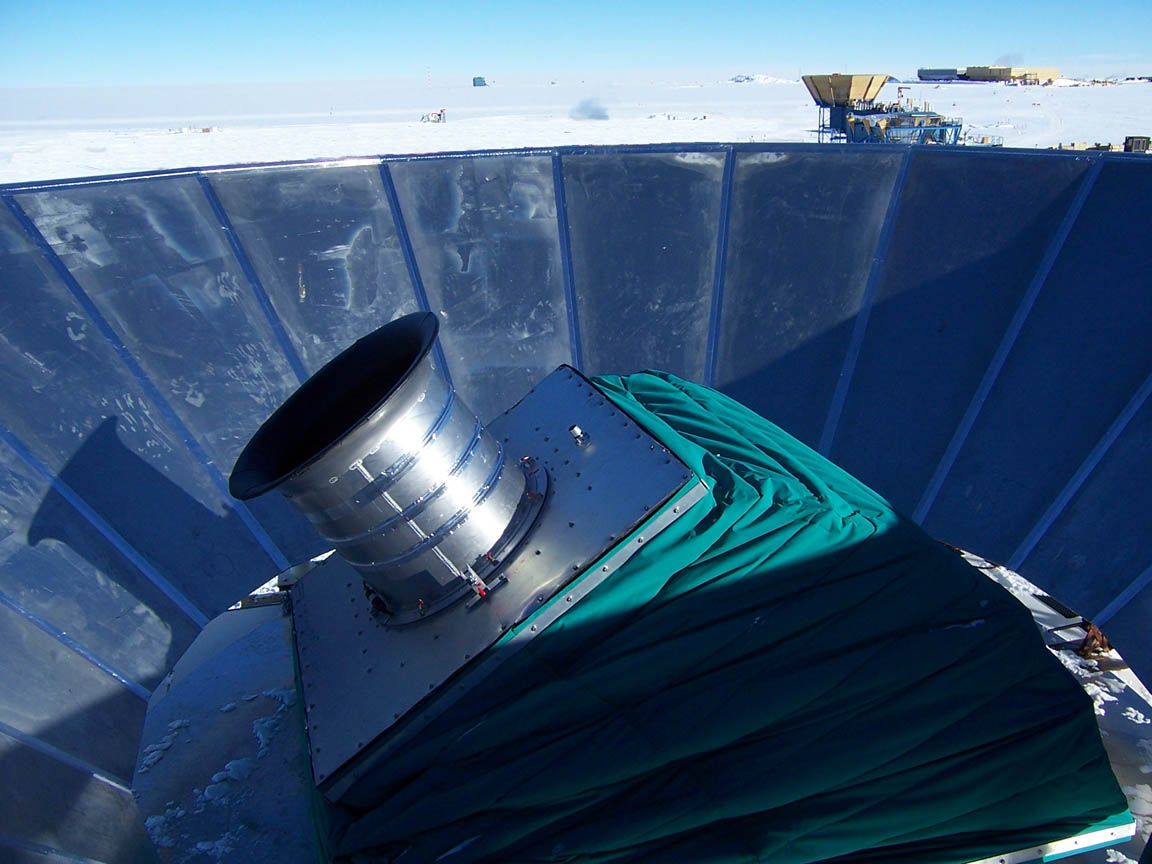}
\includegraphics[height=2.2in]{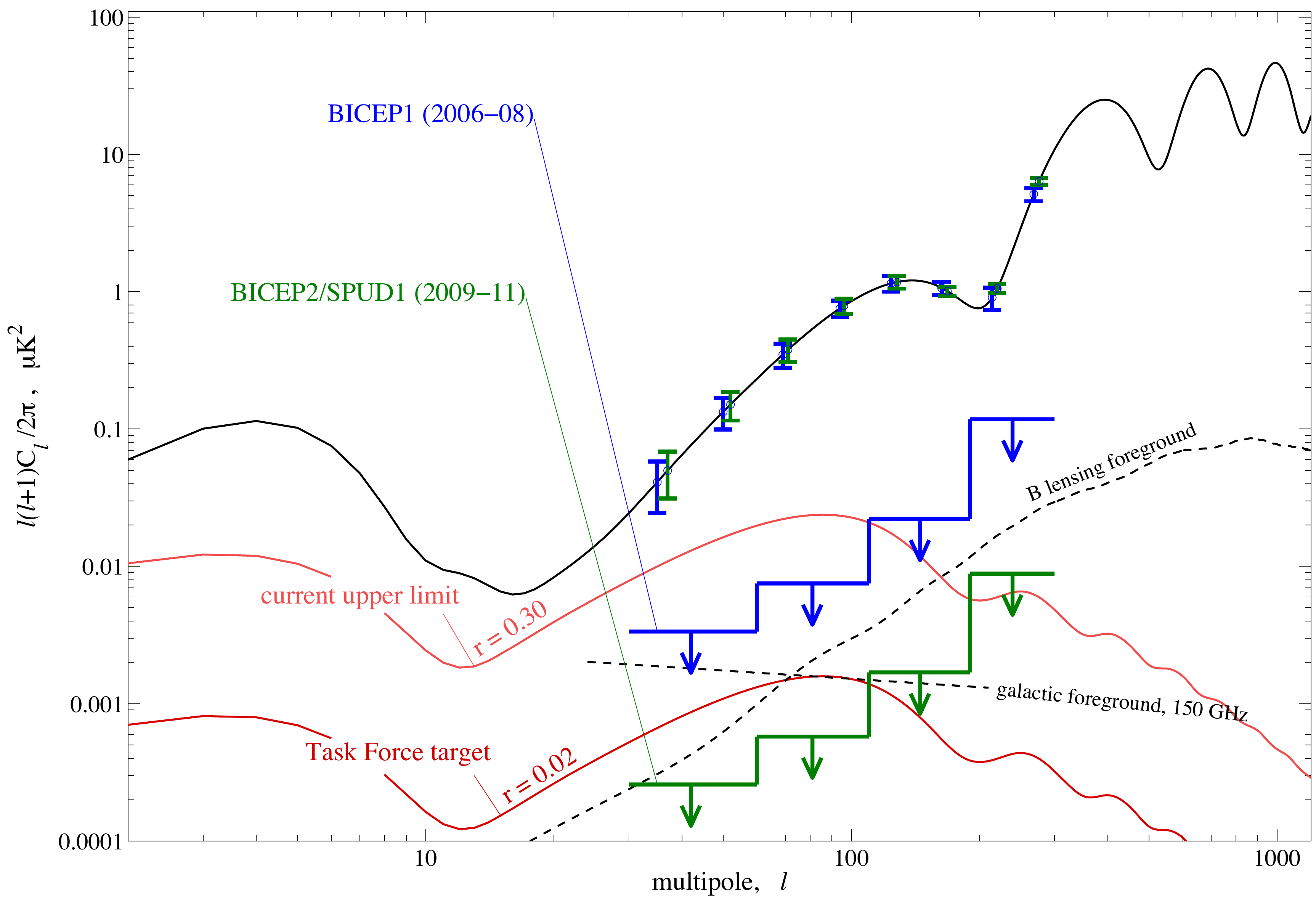} \\
\vspace{0.1in}
\includegraphics[width=0.99\textwidth]{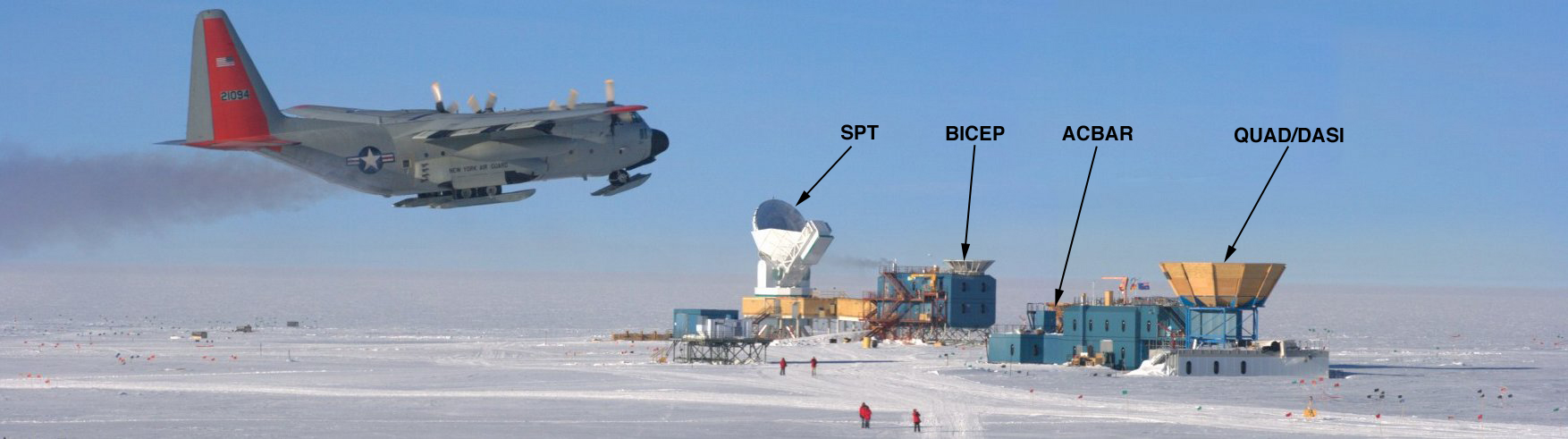}
\end{center}
\addtocounter{figure}{1}
\footnotesize
Figure 7:
The BICEP telescope (upper left) began operating from the roof of the new DSL facility in
early 2006.  With 98 PSB detectors and a small-aperture cryogenic telescope,
it is the first CMB polarimeter specifically designed to search for the signature
of gravity waves from Inflation by mapping B-mode polarization on large angular
scales (upper right).  A plan to fit upgraded small polarimeters onto
the BICEP and DASI platforms starting in 2008 (BICEP2/SPUD) will push sensitivity levels
within reach of r=0.01 Inflationary models.
The 10m South Pole Telescope (SPT) now dominates the Dark Sector skyline (lower, photo: Steffen Richter).
Visible L to R are an LC130, the SPT, DSL with the BICEP groundshield, MAPO, and
the DASI/QUAD tower and groundshield.
SPT achieved first light 16 Febuary 2007, two days before the summer's closing flight
shown here, and is now commencing an SZ survey at arcminute scales to probe the evolution of
clusters and Dark Energy.
\label{fig:dark_sector}
\end{figure}

\vspace{-0.1in}
\paragraph{BICEP and SPUD:}
The BICEP experiment, led by A. Lange at Caltech and J. Bock at JPL, is specifically optimized for
the search for degree-scale B-mode polarization from Inflationary gravity waves.
The current BICEP receiver is a sister instrument to QUAD, employing a
similar focal plane of 98 PSBs at 100 and 150~GHz.
However, its unique 30~cm aperture cryogenic refracting telescope offers the stability,
high optical throughput, and unprecedented sidelobe control critical for large
angular scale CMB polarimetry.\cite{2006SPIE.6275E..51Y}
BICEP1 operated flawlessly during its first winter;
initial results are reported in this meeting by C.~D.~Dowell.

The search for Inflationary B-modes will ultimately require increases in sensitivity
(see Figure 7) 
only achievable with vastly more detectors.
The BICEP2/SPUD project is currently developing an array of
seven monochromatic telescopes to replace BICEP1.  The first of these will be
ready for deployment on the BICEP mount in 2008 with 512 
polarization sensitive antenna-coupled TES bolometers at 150 GHz, achieving
a 9x increase in mapping speed.
Six more receivers will be ready for phased deployment onto the DASI
platform starting in 2009, promising continued sensitivity gains
without requiring new facilities or observing strategies.

\vspace{-0.1in}
\paragraph{SPT:}
The South Pole Telescope,\cite{2004SPIE.5498...11R} the product of
a large collaboration led by J. Carlstrom at U. of Chicago,
is easily the most ambitious
above-ground science facility ever built at the Pole.
Weighing 244 metric tons, the 10m off-axis Gregorian telescope
is designed to achieve 20$\mu$m surface accuracy and 1 arcsecond pointing,
specifications that should allow its eventual use in sub-mm atmospheric windows.
Its initial science goal, however, is an SZ cluster survey of up to
4000 deg$^2$.  Number counts in such a survey are sensitive to the expansion
rate and growth of structure; precision measurements can 
constrain the Dark energy equation of state.  The camera for this survey
is a 960 element TES bolometer array receiver, operating
at 90, 150, and 220 GHz.

After 14 hectic weeks of construction and assembly at Pole this summer,
SPT achieved first light in February 2007, confirming the operation of
its tracking, optics, and camera with maps of Jupiter.  Science observations
will begin this winter.  Future plans include a polarimeter receiver
which could map structure formation at high redshift with precise small angular
scale measurements of the
lensing-induced B-mode polarization of the CMB.

\section*{Acknowledgments}
We'd like to thank Francois Pajot, Bob Pernic, Steffen Richter, and Josh
Gundersen for providing some of these photos.
Scientific endeavors in the harshest place on Earth are made possible by the
National Science Foundation's Office of Polar 
Programs and the staff of the Amundsen-Scott
South Pole Station.
We thank the organizers of Rencontres du Vietnam for a fruitful conference.

\end{document}